\documentclass{aa}
\usepackage{graphicx}
\usepackage{amsmath}
\usepackage{booktabs}
\usepackage{microtype}
\usepackage{hyperref}
\usepackage{url}
\usepackage[varg]{txfonts}
\usepackage{natbib}

\newcommand{\eq}[1]{
\begin{equation}
#1
\end{equation}}
\newcommand{\st}[1]{_\text{#1}}
\newcommand{\uHz}{\hbox{\rm\thinspace $\mu$Hz}}
\newcommand{\bracfrac}[2]{\left(\frac{#1}{#2}\right)}
\newcommand{\K}{\rm\thinspace K}
\newcommand{\Msun}{\hbox{$\rm\thinspace M_{\sun}$}}
\newcommand{\Rsun}{\hbox{$\rm\thinspace R_{\sun}$}}
\newcommand{\Lsun}{\hbox{$\rm\thinspace L_{\sun}$}}
\newcommand{\yr}{\rm\thinspace yr}
\newcommand{\Myr}{\rm\thinspace Myr}
\newcommand{\Gyr}{\rm\thinspace Gyr}
\newcommand{\apm}[2]{^{+#1}_{-#2}}
\newcommand{\sci}[2]{#1\times10^{#2}}

\begin{document}

\title{A new correction of stellar oscillation frequencies for
  near-surface effects}%
\titlerunning{A new near-surface corrections for stellar oscillations}

\author{Warrick~H.~Ball\inst{1} 
  \and L.~Gizon\inst{2,1}}%
\institute{
  Institut f\"ur Astrophysik, Georg-August-Universit\"at G\"ottingen, 
  Friedrich-Hund-Platz 1, 37077 G\"ottingen, Germany\\
  \email{wball@astro.physik.uni-goettingen.de}\and 
  Max-Planck-Institut f\"ur Sonnensystemforschung, 
  Justus-von-Liebig-Weg 3, 37077 G\"ottingen, Germany}

\abstract{Space-based observations of solar-like oscillations present
  an opportunity to constrain stellar models using individual mode
  frequencies.  However, current stellar models are inaccurate near
  the surface, which introduces a systematic difference that must be
  corrected.}
{We introduce and evaluate two parametrizations of the surface
  corrections based on formulae given by Gough (1990).  The first we
  call a \emph{cubic} term proportional to $\nu^3/\mathcal{I}$ and the
  second has an additional \emph{inverse} term proportional to
  $\nu^{-1}/\mathcal{I}$, where $\nu$ and $\mathcal{I}$ are the
  frequency and inertia of an oscillation mode.}
{We first show that these formulae accurately correct model
  frequencies of two different solar models (Model S and a calibrated
  MESA model) when compared to observed BiSON frequencies.  In
  particular, even the cubic form alone fits significantly better than
  a power law. We then incorporate the parametrizations into a
  modelling pipeline that simultaneously fits the surface effects and
  the underlying stellar model parameters. We apply this pipeline to
  synthetic observations of a Sun-like stellar model, solar
  observations degraded to typical asteroseismic uncertainties, and
  observations of the well-studied CoRoT target HD~52265.  For
  comparison, we also run the pipeline with the scaled power-law
  correction proposed by Kjeldsen et al. (2008).}
{The fits to synthetic and degraded solar data show that the method is
  unbiased and produces best-fit parameters that are consistent with
  the input models and known parameters of the Sun.  Our results for
  HD~52265 are consistent with previous modelling efforts and the
  magnitude of the surface correction is similar to that of the Sun.
  The fit using a scaled power-law correction is significantly worse
  but yields consistent parameters, suggesting that HD~52265 is sufficiently
  Sun-like for the same power-law to be applicable.}
{We find that the cubic term alone is suitable for asteroseismic
  applications and it is easy to implement in an existing pipeline.
  It reproduces the frequency dependence of the surface correction
  better than a power-law fit, both when comparing calibrated solar
  models to BiSON observations and when fitting stellar models using
  the individual frequencies.  This parametrization is thus a useful
  new way to correct model frequencies so that observations of
  individual mode frequencies can be exploited.}


\maketitle

\section{Introduction}

The {\it Kepler} \citep{kepler} and CoRoT \citep{corot} missions have
ushered in a new era for asteroseismology.  Hundreds of main-sequence
and subgiant stars have now been observed at sufficiently short
cadence to resolve their solar-like oscillations.  Of these,
individual oscillation frequencies have been identified in dozens of
stars \citep[e.g.][]{appourchaux2012} for which many have
complementary spectroscopic observations
\citep[e.g][]{bruntt2012,molenda2013} and a handful additionally have
interferometric constraints on their radii
\citep{huber2012,white2013}.  This wealth of observational data
presents the opportunity to better constrain stellar model parameters,
as well as the physics of the models themselves.

The main obstruction to constraining stellar models using the
individually-identified frequencies is the known systematic difference
between models and observations caused by improper modelling of the
near-surface layers.  The frequency shifts are produced by several
neglected or poorly-modelled physical processes and are collectively
known as the \emph{surface effects} or \emph{surface term}.  Specific
examples include poor modelling of temperature gradients in the
superadiabatic layer, the use of the adiabatic approximation when
calculating oscillation frequencies, and the absence of a description
of interactions between convection and the oscillations.

\begin{figure}
\includegraphics[width=85mm]{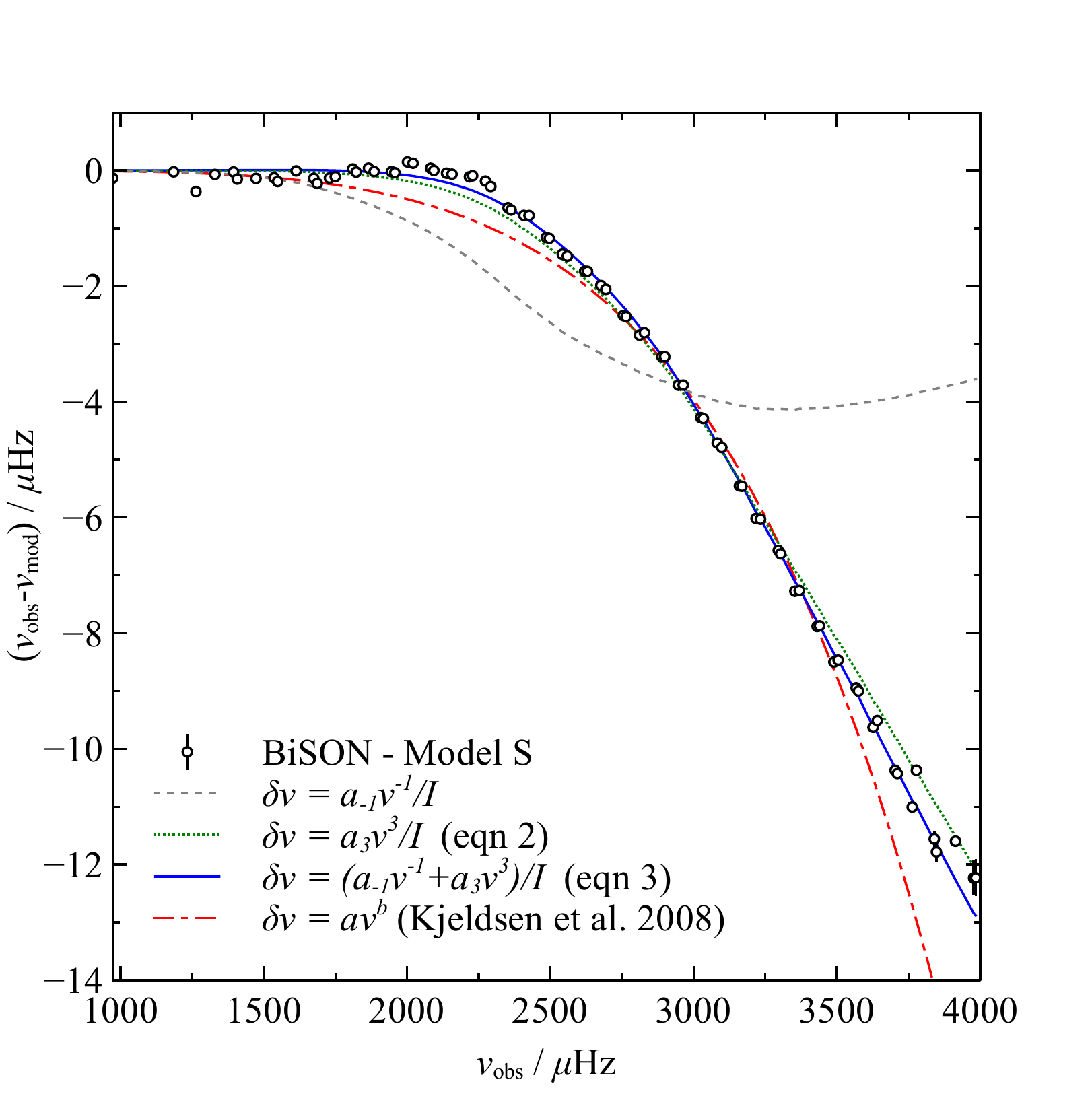}
\caption{Frequency differences between a standard solar model
  \citep[Model S,][]{modelS} and observations of low-degree modes
  ($\ell\leq3$) by BiSON.  The lines show fits made using an inverse
  term (dashed), cubic term (dotted) or both terms (solid).  The fit
  with the inverse term is quite poor, with the cubic term much better
  and with both terms somewhat better still.  The dot-dashed lines
  show a power law, fit to nine radial orders about
  $\nu\st{max}=3090\uHz$, as is used in the frequency correction
  proposed by \citet{kbcd2008}.}
\label{modelS_BiSON}
\end{figure}

Several methods have already been employed to reduce the bias induced
by these surface effects on the underlying parameters of stellar
models.  First, \citet{kbcd2008} proposed modelling the surface
effects as a power-law in frequency, calibrated to radial oscillations
in the Sun around the frequency of maximum oscillation power
$\nu\st{max}$, and fit to the differences between observed and
modelled frequencies, the latter rescaled by the mean stellar density
relative to the Sun.  This scaled power-law approach has been adopted
widely \citep[e.g.][]{tang2011, mathur2012, deheuvels2012}.  However,
the fit overpredicts the magnitude of the surface effect at lower
frequencies (see Figs~\ref{modelS_BiSON} and \ref{modelM_BiSON}), and
the correction assumes that the frequency shifts follow the same power
law in other stars.

Second, \citet{roxburgh2003} suggested fitting models to observations
using particular ratios of frequency differences, which have been used
by several groups \citep[e.g.][]{silva2013} to fit stellar models.  By
approximating the eigenfunctions with partial wave solutions, one can
express the surface effect as a phase shift that can be suppressed by
choosing appropriate ratios of frequency differences.  \citet{oti2005}
computed the structural sensitivity kernels for these ratios and
demonstrated that they are indeed more sensitive to the stellar core.
Because this method uses differences and ratios, there is some loss of
information, but the advantage is that stellar model parameters should
not be biased by uncertainty at the surface.

Third, \citet{ggk1} proposed a Bayesian method in which an additional
frequency offset parameter is introduced for each oscillation mode and
then marginalized against some prior.  They used their method to
compare the statistical evidence for solar models with different input
physics \citep{ggk2}, and to model a number of Kepler targets
\citep{ggk3}.  The method appears successful in solar modelling, but,
as noted by \citet{ggk1} in their closing sections, the results appear
biased towards larger masses when too few low-frequency modes are
available.

Motivated by a lack of a leading method for modelling surface effects,
we propose here a new method in which the surface effects are modelled
by one or both of terms proportional to $\nu^{-1}/\mathcal{I}$ and
$\nu^3/\mathcal{I}$, where $\nu$ is the frequency of an oscillation
mode and $\mathcal{I}$ its corresponding inertia, normalized by the
total displacement at the photosphere.  In Section 2, we motivate
these parametrizations and demonstrate the quality of fits to the
differences between modelled and observed low-degree oscillations in
the Sun.  In Section 3, we test the robustness of the parametrizations
against both synthetic and real solar data.  In Section 4, we apply
the method to the planet-hosting CoRoT target HD~52265, with promising
results, given the quality of the fit and consistency with previous
results.  Finally, we discuss the performance of the method and its
potential flaws.

\section{Modelling and fitting of surface effects}

\subsection{Parametrizations}

\begin{figure}
\includegraphics[width=85mm]{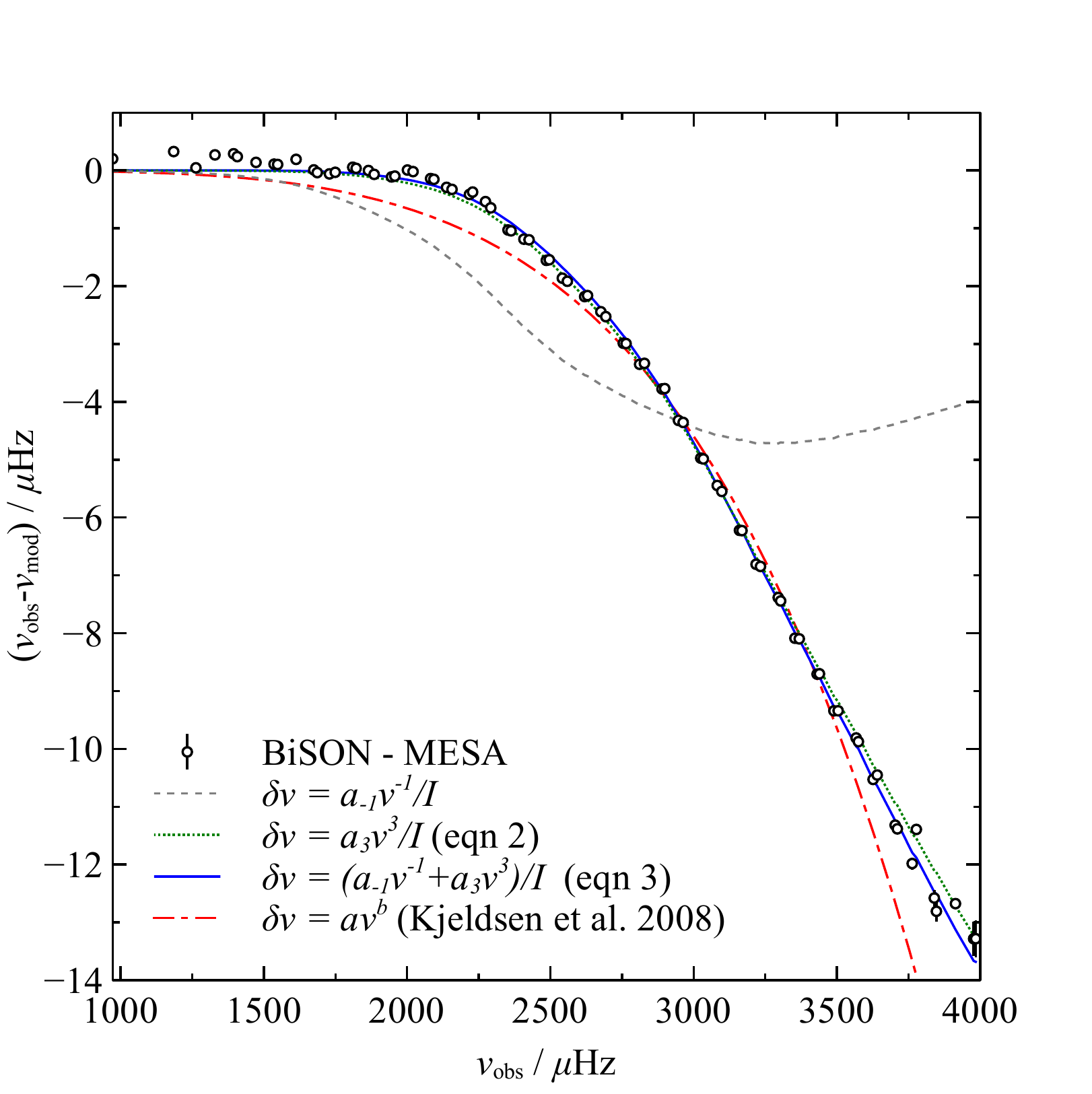}
\caption{Frequency differences between a solar model calibrated with
  MESA and observations of low-degree modes ($\ell\leq3$) by BiSON.
  The lines show fits made using an inverse term (dashed), cubic term
  (dotted) or both terms (solid).  The relative performances of the
  fits is the same as Fig.~\ref{modelS_BiSON}.  The dot-dashed lines
  show a power law, fit to nine radial orders about
  $\nu\st{max}=3090\uHz$, as is used in the frequency correction
  proposed by \citet{kbcd2008}.}
\label{modelM_BiSON}
\end{figure}

\citet{gough1990} discussed potential asymptotic forms for the
frequency shifts observed by \citet{libbrecht1990} over the solar
activity cycle.  By considering perturbations near the surface under a
variational principle and taking the asymptotic form of the
displacement eigenfunctions in the evanescent layer,
\citet{gough1990} concluded that the frequency shifts should generally
take the form \citep[equation 9.3 of ][]{gough1990}
\eq{\delta\nu\propto\frac{Q(\nu^2)}{\nu \mathcal{I}}\text{,}} %
where $Q(x)$ is a quadratic function in $x$; $\nu$ is the cyclic
frequency of an oscillation mode; and $\delta\nu$ is the frequency
shift induced by a perturbation near the stellar surface.  The
normalized mode inertia $\mathcal{I}$ is defined by 
\citep[e.g.][equation 3.140]{acdk2010}%
\eq{\mathcal{I}=
  \frac{4\pi\int_0^R\left[|\xi\st{r}(r)|^2+\ell(\ell+1)|\xi\st{h}(r)|^2\right]\rho
    r^2dr}
  {M\left[|\xi\st{r}(R)|^2+\ell(\ell+1)|\xi\st{h}(R)|^2\right]}\text{,}}
where $\xi\st{r}$ and $\xi\st{h}$ are the radial and horizontal
components of the displacement eigenvector, $R$ the photospheric
radius, $\rho$ the unperturbed stellar density, $M$ the total stellar
mass, and $\ell$ the degree of the mode.  In the solar models, the
mode inertia decreases rapidly with frequency below about $2000\uHz$
before levelling out and reaching a minimum around $4000\uHz$.  This
behaviour suppresses the magnitude of the frequency shifts at low
frequency.

\citet{gough1990} further argued that a perturbation caused by a
magnetic field concentrated into filaments, which would mostly modify
the sound speed without much affecting the gas density, would cause a
shift proportional to $\nu^3/\mathcal{I}$.  This is the same form
suggested by \citet{goldreich1991} for photospheric perturbations.  A
perturbation caused by a change in the description of convection,
which would presumably modify the pressure scale height, would cause a
shift proportional to $\nu^{-1}/\mathcal{I}$.  Following from their
distinct dependences on frequency, we refer to these two terms as
\emph{cubic} and \emph{inverse} surface effects, and to their
combination as a \emph{combined} surface effect.

Fig.~\ref{modelS_BiSON} shows the difference between low-degree
($\ell\leq3$) solar oscillation frequencies\footnote{We have used the
  `quiet-Sun' frequencies given in Table 3 of \citet{broomhall2009}.}
observed over 8640 days by the Birmingham Solar-Oscillations Network
\citep[BiSON,][]{broomhall2009} and adiabatic oscillation frequencies
computed for a standard solar \citep[Model S,][]{modelS}.  The curves
represent uncertainty-weighted least-squares fits to the observed
frequency differences using either the cubic term, the inverse term or
both.  The inverse term alone does not produce a good fit to the
differences ($\chi^2=424\,092$), whereas the cubic term does
($\chi^2=16\,150$).  The combination of both terms provides a
significantly better fit ($\chi^2=11\,250$).  Similarly,
Fig.~\ref{modelM_BiSON} shows the difference between the BiSON
frequencies and a solar model computed using the solar calibration
test case distributed with the Modules for Experiments in Stellar
Astrophysics
\citep[MESA\footnote{\url{http://mesa.sourceforge.net/}},][]{paxton2011,paxton2013}.
Again, the inverse term alone fits badly ($\chi^2=517\,678$), the
cubic term better ($\chi^2=27\,287$), and their combination slightly
better still ($\chi^2=25\,728$).  The larger $\chi^2$ of the MESA
solar model is chiefly contributed by a handful of modes below
$1500\uHz$, where the MESA model shows systematically higher
frequencies and the observed uncertainties are smallest.  In fact, the
two lowest-frequency modes alone contribute about $17\,000$ to
$\chi^2$.  Despite this difference, both solar models fit best with
similar coefficients for the surface terms.

Figs~\ref{modelS_BiSON} and \ref{modelM_BiSON} also show power-law
fits as suggested by \citet{kbcd2008}.  For each of the solar models,
we fit a power law to the frequency differences for nine radial orders
about $3090\uHz$.  Though the fits are reasonably good in the fitted
range, they are distinctly worse at higher and lower frequencies, as
can be seen by the overprediction of the size of the correction around
$2200\uHz$.  When one considers all the modes, the power-law fits give
$\chi^2=49\,981$ and $90\,657$ for Model S and the MESA solar model,
or $46\,626$ and $75\,172$ when also fit to all the observed
frequencies.  The deviation is largely a result of the change of the
frequency dependence of the mode inertiae around the $2500\uHz$.
Thus, when restricted to frequencies around $3090\uHz$, the fit is
comparably good, but its accuracy declines as the frequency range
increases.

We thus propose two potential parametrizations for surface effects.
Firstly, we suggest a purely cubic term,
\eq{\delta\nu=a_3(\nu/\nu\st{ac})^3/\mathcal{I}\text{,}\label{form1}}
or, secondly, a combined expression,
\eq{\delta\nu=\left(a_{-1}(\nu/\nu\st{ac})^{-1}
    +a_3(\nu/\nu\st{ac})^3\right)/\mathcal{I}\text{,}\label{form2}}
where $a_{-1}$ and $a_3$ are coefficients that are to be fit for the
stellar model under consideration, and $\nu\st{ac}$ is the acoustic
cutoff frequency, used here purely to non-dimensionalize the
frequencies.  For convenience, we compute $\nu\st{ac}$ using the
scaling relation
\eq{\nu\st{ac}/\nu_{\text{ac},\odot}=\frac{g}{g_\odot}\bracfrac{T\st{eff}}{T_{\text{eff},\odot}}^{-1/2}}
where we take $\nu_{\text{ac},\odot}=5000\uHz$.  

\subsection{Fitting the surface effect}

This parametrization uses only global parameters of the oscillations
(i.e. frequency $\nu$ and mode inertia $\mathcal{I}$), both of which
are typically computed by oscillation packages
\citep[e.g. ADIPLS,][]{adipls}.  In addition, the parametrization only
introduces linear terms, which can be fit analytically.  It is not
necessary to construct any additional stellar models or use iterative
methods to compute the coefficients $a_{-1}$ and $a_3$, so the fit can
easily be incorporated into existing model-fitting pipelines.  We now
describe in detail how this can be performed.

Suppose that we are comparing a stellar model, whose mode frequencies
$\nu_{i,\text{mod}}$ and inertiae $\mathcal{I}_i$ have been computed, to
a set of observed mode frequencies $\nu_{i,\text{obs}}$ with
uncertainties $\sigma_i$.  Here, $i\in\{1,\ldots,N\st{modes}\}$,
where $N\st{modes}$ is the number of observed modes.  The best-fitting
coefficients for the surface effect are determined by minimizing%
\eq{\chi^2\st{SE}=\sum_{i=1}^{N\st{modes}}
  \bracfrac{\nu_{i,\text{mod}}+\delta\nu_i-\nu_{i,\text{obs}}}{\sigma_i}^2}
with respect to the coefficients $a_3$ and $a_{-1}$, with
$\delta\nu_i$ given by either of equations \ref{form1} or \ref{form2},
above.  Note that the coefficients are optimized with respect to all
the observed frequencies rather than just the radial modes, as in the
formula of \citet{kbcd2008}.

Because the surface effect is linear in the coefficients, the
best-fitting values can be determined analytically by linear
regression.  Let us first define the vector $\vec{y}$ by%
\eq{y_i=\frac{\nu_{i,\text{obs}}-\nu_{i,\text{mod}}}{\sigma_i}\text{.}}%
For the cubic surface effect (equation \ref{form1}), we further define
the vector $\vec{X}$ by%
\eq{X_i=\frac{\nu_{i,\text{mod}}^3}{\mathcal{I}_i\sigma_i}\text{,}}%
in which case the best-fitting coefficient $a_3$ is %
\eq{a_3=\frac{\sum_i{X_iy_i}}{\sum_i{X_i^2}}\text{.}}
For the combined surface effect (equation \ref{form2}), we instead
define a matrix $\tens{X}$ by%
\eq{X_{i,1}=\frac{\nu_{i,\text{mod}}^{-1}}{\mathcal{I}_i\sigma_i}\text{,}\quad
  X_{i,2}=\frac{\nu_{i,\text{mod}}^3}{\mathcal{I}_i\sigma_i}\text{,}}%
in which case the best-fitting coefficients $a_{-1}$ and $a_3$ are
given by%
\eq{\left(\begin{array}{c}a_{-1} \\ a_3 \end{array}\right)=
  \left(\tens{X}^\text{T}\tens{X}\right)^{-1}\tens{X}^\text{T}\vec{y}\text{.}}%
This matrix equation does not reduce to as simple a form as the cubic
case but the equations are still easily solved analytically.  The
matrix $\tens{X}^\text{T}\tens{X}$ is only $2\times2$-dimensional, so
its inverse is trivially computed.
Note that $(\tens{X}^\text{T}\tens{X})^{-1}\tens{X}^\text{T}$ is simply the
Moore--Penrose pseudoinverse of $\tens{X}$.

Thus, given the mode frequencies and inertiae of a stellar model, the
analytic calculation described above returns the optimal values of the
coefficients in a least-squares sense.  Unlike the method of
\citet{kbcd2008}, the frequencies are not rescaled by the ratio of the
modelled and observed large separation before being corrected.  In
their formulation, this constrains the overall scale of the frequency
correction to the same as those models that have a similar large
separation.  Without such a rescaling, it is possible that an
incorrect model could be found to fit well with unrealistically large
values of the coefficients.  In fact, in the new parametrization
presented here, even the sign of the surface correction is currently
not constrained, so there is nothing to prevent a best-fitting model
that has a surface effect of opposite sign to that of the Sun.  We
have not found this to be a problem but we cannot demonstrate that it
never will be.  

A possible solution of this issue would be to penalize large values of
the coefficients by choosing an appropriate prior.  For example, one
could assume that the logarithms of the coefficients are uniformly
distributed, which is often done for parameters that represent a scale
rather than a position in parameter space.  This would also prevent
values of opposite sign but would require modification of the analytic
calculation above.

\subsection{Stellar models and fitting procedure}

We tested the parametrizations, using the method outline above, with
the downhill simplex method \citep{nelder65} implemented in MESA
(revision 6022) optimizes the values of a stellar model's age $t$,
mass $M$, initial metallicity $[$Fe/H$]_i$, initial helium abundance
$Y_i$ and mixing-length parameter $\alpha$.  The combined
spectroscopic constraints and corrected frequencies were used to
evaluate the objective function given by
\eq{\chi^2=\sum_{i=1}^N\bracfrac{x_{i,\text{mod}}-x_{i,\text{obs}}}{\sigma_i}^2\text{,}}
where $x_{i,\text{obs}}$ represent all of the $N$ observations
used--spectroscopic or seismic--, $\sigma_i$ their uncertainties, and
$x_{i,\text{mod}}$ the corresponding model values.  That is, we do not
distinguish between any of the parameters in computing $\chi^2$, as in
e.g.~\citet{metcalfe2014}.

For comparison, we also ran the pipeline using the surface correction
proposed by \citet{kbcd2008}, in which the surface correction is
modelled as a power-law in frequency, calibrated to the Sun.  By
fitting the frequency differences between the BiSON data and the MESA
solar model above for nine radial orders about $\nu=3090\uHz$, we
determined the appropriate index to be $b=4.81$.  This is similar to
the value $b=4.9$ reported by \citet{kbcd2008}, which is often used
without recalibration.

To compute an initial model for each evolutionary track, we first
initialized a pre-main-sequence model with a central temperature
$300\,000\K$ and allowed it to evolve until $\log L/\Lsun=0.4$,
where $L$ and $\Lsun$ are the total stellar and solar luminosities.
At the beginning of each track, this model was then rescaled by mass
to the value being evaluated, from its existing mass of $1\Msun$ for
the synthetic and degraded solar fits, and $1.2\Msun$ for HD~52265.
In both cases, these models had ages less than $2\Myr$.  Stellar
models were evolved starting from the pre-main-sequence model
described above until a minimum of $\chi^2$ had been found and one of
the spectroscopic constraints was beyond its observed value by
$5\sigma$.  The parameters of the stellar model (including the age of
the optimal fit along that track) are then used to update the simplex.

For the constitutive physics, we use opacities from the OPAL tables
\citep{iglesias1996} and \citet{ferguson2005} at high and low
temperatures, respectively, blended linearly in the range
$4.00\leq\log T\leq4.10$.  A standard Eddington grey atmosphere is
employed and convective processes are described by mixing-length
theory according to \citet{henyey1965}.  In the synthetic and
Sun-as-a-star fits, convective overshooting was fixed at its
solar-calibrated value.  For HD~52265, no overshooting was included.
The composition was scaled from the solar values given by
\citet{grevesse1998}.  Diffusion and gravitational settling are
included through the method of \citet{thoul1994}.  Nuclear reaction
rates are drawn from \citet{caughlan1988} or the NACRE collaboration
\citep{angulo1999}, with preference given to the latter when
available.  We used newer specific reaction rates for the reactions
${}^{14}\mathrm{N}(p,\gamma){}^{15}\mathrm{O}$ \citep{imbriani2005}
and ${}^{12}\mathrm{C}(\alpha,\gamma){}^{16}\mathrm{O}$
\citep{kunz2002}.
All other options took default values,
which are described in the MESA instrument papers
\citep{paxton2011,paxton2013} and available in the source code.

The oscillation frequencies and mode inertiae were computed using the
Aarhus adiabatic oscillation package \citep[ADIPLS,][]{adipls},
without remeshing the stellar model.  The models were computed with a
mesh-spacing coefficient $C=1/2$, which corresponds to about 1500
meshpoints.  This is sufficient to correctly compute oscillation
frequencies and mode inertiae along the main sequence.  Calls to
ADIPLS are made from MESA during the stellar evolution calculation and
by default return the frequencies and inertiae.

We set the parameter tolerance in the downhill simplex method to
$10^{-5}$.  
For the synthetic data, the initial parameters used the same
parameters as for the input model.  For the Sun, the initial
parameters were the calibrated values given by \citet{paxton2013}.
Finally, for HD~52265, we began with parameters computed by a fit to
the spectroscopic parameters and averaged large and small separations,
as measured by \citet{ballot2011}, and then found best-fit model
parameters for the unperturbed
observations with a cubic surface effect.  These parameters were used
as the initial guess for all the fits to HD~52265.

We then used the observed uncertainties to generate 100 random
realizations of the observations and fit a stellar model to each
realization, using the parameters of the primary model as the initial
guess.  The reported values are the $15.85$, $50$ and $84.15$
percentiles for the relevant distributions. i.e. the median and a
$68.3$ per cent confidence interval for the sample of 100 fits.

Because different choices of timestep can lead to slightly different
models for a given set of parameters, the timesteps were fixed at
$10\Myr$ when fitting the synthetic data, the same value used to
compute the input model.  For the fits to the degraded BiSON
observations and HD~52265, the timestep was decreased adaptively from
$30\Myr$ to $300\,000\yr$ based on partial evaluations of $\chi^2$
dropping below certain limits.

The results of all of the fits are presented in Table~\ref{supertable}.

\begin{sidewaystable*}
\tiny
\caption{Results of all fits presented here, as well as fits for
  HD~52265 presented by \citet{escobar2012} and spectroscopic inputs for
  all fits (labelled \emph{Input}).  The central values are the
  medians and the upper and lower limits are the locations of the
  $15.85$ and $85.15$ percentiles.  In the case of the synthetic data,
  we also include the parameter values used to generate the
  observations.}
\label{supertable}
\begin{tabular}{ccccccccccccc}
\toprule
Synthetic & $t/\text{Gyr}$ & $M/\Msun$ & $[$Fe/H$]_i$ & $Y_i$ & $\alpha$ & 
$a_3/10^{-7}\uHz$ & $a_{-1}/10^{-9}\uHz$ &
$T\st{eff}/\text{K}$ & $\log g$ & $\log L/\Lsun$ & $[$Fe/H$]_s$ &
$R/\Rsun$ \\
\midrule
Input & $4.61$ & $1.000$ & $0.097$ & $0.2784$ & $1.908$ &
& &
$5773\pm60$ & $4.45\pm0.09$ & $-0.01\pm0.05$ & $0.04\pm0.05$ &
$0.9918$ \\
A & $4.64\apm{0.23}{0.27}$ & $0.999\apm{0.002}{0.005}$ & 
$0.098\apm{0.011}{0.014}$ & $0.2780\apm{0.0006}{0.0004}$ & 
$1.90\apm{0.03}{0.05}$ & 
$0.00\apm{0.06}{0.10}$ & &
$5766\apm{43}{41}$ & $4.4451\apm{0.0006}{0.0008}$ & 
$-0.010\apm{0.012}{0.014}$ & $0.040\apm{0.012}{0.014}$ & 
$0.9911\apm{0.0006}{0.0016}$ \\
B & $4.58\apm{0.26}{0.21}$ & $0.999\apm{0.002}{0.005}$ & 
$0.096\apm{0.014}{0.020}$ & $0.2781\apm{0.0005}{0.0004}$ & 
$1.91\apm{0.03}{0.06}$ & 
$-1.90\apm{0.10}{0.08}$ & & 
$5769\apm{46}{54}$ & $4.4452\apm{0.0004}{0.0009}$ & 
$-0.010\apm{0.014}{0.017}$ & $0.038\apm{0.014}{0.020}$ & 
$0.9911\apm{0.0006}{0.0014}$ \\
C & $4.64\apm{0.28}{0.25}$ & $0.999\apm{0.012}{0.009}$ & 
$0.093\apm{0.037}{0.019}$ & $0.2778\apm{0.0009}{0.0011}$ & 
$1.91\apm{0.08}{0.09}$ & 
$-0.01\apm{0.23}{0.20}$ & $0.71\apm{6.90}{7.88}$ &
$5772\apm{70}{78}$ & $4.4451\apm{0.0016}{0.0015}$ & 
$-0.009\apm{0.023}{0.026}$ & $0.035\apm{0.039}{0.019}$ & 
$0.9911\apm{0.0041}{0.0032}$ \\
D & $4.62\apm{0.31}{0.23}$ & $1.000\apm{0.015}{0.011}$ & 
$0.093\apm{0.037}{0.033}$ & $0.2778\apm{0.0011}{0.0015}$ & 
$1.91\apm{0.11}{0.07}$ & 
$-2.05\apm{0.32}{0.24}$ & $4.22\apm{6.77}{6.37}$ & 
$5768\apm{78}{52}$ & $4.4453\apm{0.0017}{0.0019}$ & 
$-0.011\apm{0.029}{0.018}$ & $0.034\apm{0.040}{0.035}$ & 
$0.9913\apm{0.0053}{0.0035}$ \\
\midrule
\multicolumn{3}{l}{Sun-as-a-star} \\
\midrule
Input & & & & & & & & 
$5778\pm60$ & $4.44\pm0.09$ & $0.000\pm0.024$ & $0.00\pm0.05$ \\
E & $4.74\apm{0.18}{0.19}$ & $1.000\apm{0.015}{0.010}$ & 
$0.083\apm{0.020}{0.045}$ & $0.276\apm{0.006}{0.018}$ & 
$1.83\apm{0.05}{0.07}$ & 
$-2.47\apm{0.14}{0.44}$ & & 
$5768\apm{48}{46}$ & $4.4382\apm{0.0020}{0.0013}$ & 
$-0.001\apm{0.014}{0.017}$ & $0.023\apm{0.022}{0.047}$ & 
$1.000\apm{0.005}{0.003}$ \\
F & $4.68\apm{0.30}{0.19}$ & $1.000\apm{0.014}{0.007}$ & 
$0.085\apm{0.017}{0.043}$ & $0.278\apm{0.006}{0.017}$ & 
$1.84\apm{0.04}{0.05}$ & 
$-2.45\apm{0.15}{0.26}$ & $2.34\apm{7.62}{7.61}$ & 
$5779\apm{38}{43}$ & $4.4383\apm{0.0020}{0.0013}$ & 
$0.002\apm{0.012}{0.014}$ & $0.026\apm{0.017}{0.047}$ & 
$1.000\apm{0.004}{0.003}$ \\
G & $4.75\apm{0.25}{0.32}$ & $1.003\apm{0.010}{0.007}$ & 
$0.091\apm{0.012}{0.012}$ & $0.277\apm{0.005}{0.011}$ & 
$1.87\apm{0.04}{0.05}$ & & &
$5786\apm{58}{50}$ & $4.4381\apm{0.0015}{0.0009}$ & 
$0.007\apm{0.014}{0.014}$ & $0.033\apm{0.013}{0.012}$ & 
$1.001\apm{0.003}{0.002}$ \\
\midrule
\multicolumn{3}{l}{HD~52265} \\
\midrule
Input & & & & & & & & 
$6100\pm60$ & $4.35\pm0.09$ & $0.328\pm0.024$ & $0.19\pm0.05$ \\
H & $2.44\apm{0.16}{0.15}$ & $1.215\apm{0.030}{0.008}$ & 
$0.259\apm{0.030}{0.040}$ & $0.292\apm{0.006}{0.029}$ & 
$1.64\apm{0.07}{0.06}$ & 
$-3.40\apm{0.49}{0.51}$ & &
$6056\apm{40}{40}$ & $4.2853\apm{0.0052}{0.0012}$ & 
$0.322\apm{0.015}{0.013}$ & $0.161\apm{0.034}{0.042}$ & 
$1.314\apm{0.008}{0.002}$ \\
I & $2.44\apm{0.13}{0.09}$ & $1.211\apm{0.007}{0.007}$ & 
$0.264\apm{0.029}{0.026}$ & $0.294\apm{0.005}{0.004}$ & 
$1.63\apm{0.07}{0.04}$ & 
$-3.41\apm{0.39}{0.36}$ & $11.19\apm{7.39}{5.77}$ &
$6057\apm{55}{38}$ & $4.2842\apm{0.0009}{0.0013}$ & 
$0.320\apm{0.014}{0.010}$ & $0.164\apm{0.033}{0.029}$ & 
$1.313\apm{0.004}{0.003}$ \\
J & $2.26\apm{0.18}{0.11}$ & $1.222\apm{0.020}{0.005}$ & 
$0.312\apm{0.024}{0.035}$ & $0.301\apm{0.006}{0.009}$ & 
$1.66\apm{0.08}{0.06}$ & & &
$6073\apm{56}{58}$ & $4.2854\apm{0.0016}{0.0010}$ & 
$0.329\apm{0.018}{0.012}$ & $0.218\apm{0.030}{0.039}$ & 
$1.319\apm{0.008}{0.003}$ \\
TGEC & $2.6\pm0.2$ & $1.24\pm0.02$ & $0.27\pm0.04$ & $0.28\pm0.02$ & &
& &
$6120\pm20$ & $4.284\pm0.002$ & $0.348\pm0.006$ & $0.20\pm0.04$ & $1.33\pm0.02$ \\
AMP(a) & $3.00$ & $1.22$ & $0.23$ & $0.280$ & & & &
6019 & 4.282 & 0.313 & & 1.321 \\
AMP(b) & $2.38$ & $1.20$ & $0.215$ & $0.298$ & & & &
6097 & 4.282 & 0.328 & & 1.310 \\
\bottomrule
\end{tabular}
\end{sidewaystable*}

\section{Tests on synthetic and solar data}

\subsection{Synthetic data}

We first tested whether our method correctly recovers synthetic input
data.  We constructed a stellar model with input parameters taken from
the solar calibration presented by \citet{paxton2013} and computed its
oscillation frequencies and mode inertiae.  We then used its
observable properties as inputs for our modelling pipeline, using the
same selection of observables as is available for HD~52265, along with
the corresponding uncertainties.  The spectroscopic parameters are the
effective temperature $T\st{eff}$, surface gravity $\log g$,
metallicity $[$Fe/H$]_s$ and luminosity $\log L/\Lsun$.  We fit the
frequencies of 28 oscillation modes: 
10 radial, 10 dipole, and 8 quadrupole modes, centred on
$\nu\st{max}\approx3000$.  These observable constraints are fairly
typical for a star with good spectroscopic and intermediate seismic
observations, with a timeseries of 117 days.

We first tested the cubic-only surface effect, first by fitting the
synthetic data without any surface effect (case A) and second by
fitting with a surface effect with $a_3=\sci{-1.87}{-7}\uHz$ (case B).
We then tested the combined surface effect, again first without a
surface effect (case C) and then with an artificial surface effect
(case D) with $a_{-1}=\sci{2.07}{-7}\uHz$ and
$a_3=\sci{-2.78}{-9}\uHz$.  The results are labelled correspondingly
in Table~\ref{supertable}.

In all four cases, the fitted parameters are consistent with the input
parameters, be it with or without a surface effect, cubic or combined.
We also confirmed that the parameters of the input model give a
perfect fit.~i.e. $\chi^2=0$, as expected.

In cases C and D, $a_{-1}$ is consistent with the input value but
comes with large uncertainties.  In fact, in case D, the result is
consistent with $a_{-1}=0$.  This is unsurprising, given that the
range of observed frequencies does not extend to the bend in the
surface effect caused by the sharp frequency dependence of the mode
inertiae at low frequencies.  In addition, Figs~\ref{modelS_BiSON} and
\ref{modelM_BiSON} show that the surface effect is dominated by the
cubic term, especially at frequencies about about $2500\uHz$.

\begin{figure}
\includegraphics[width=85mm]{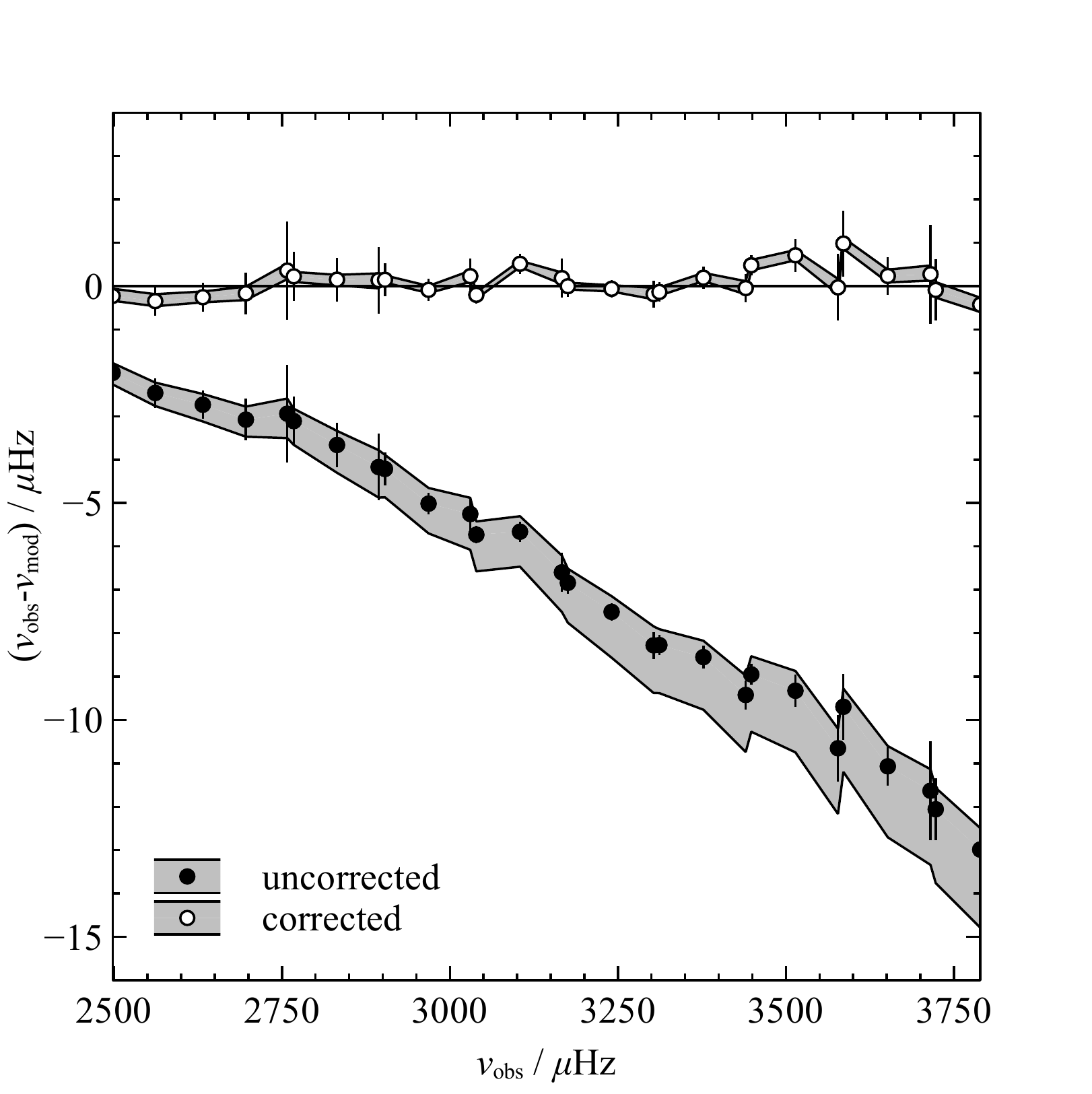}
\caption{Uncorrected and corrected frequency differences between
  observed frequencies for the degraded solar data and MESA models fit
  using only the cubic term (equation \ref{form1}).  The solid and
  empty points show the uncorrected and corrected differences with
  error bars that correspond to the observed uncertainties.  The
  shaded regions show the spread of the modelled
  frequencies. i.e. taken from the 100 fits to random realizations of
  the observations.  There is no obvious remaining trend in the
  corrected frequencies.}
\label{saas_cube}
\end{figure}

\begin{figure}
\includegraphics[width=85mm]{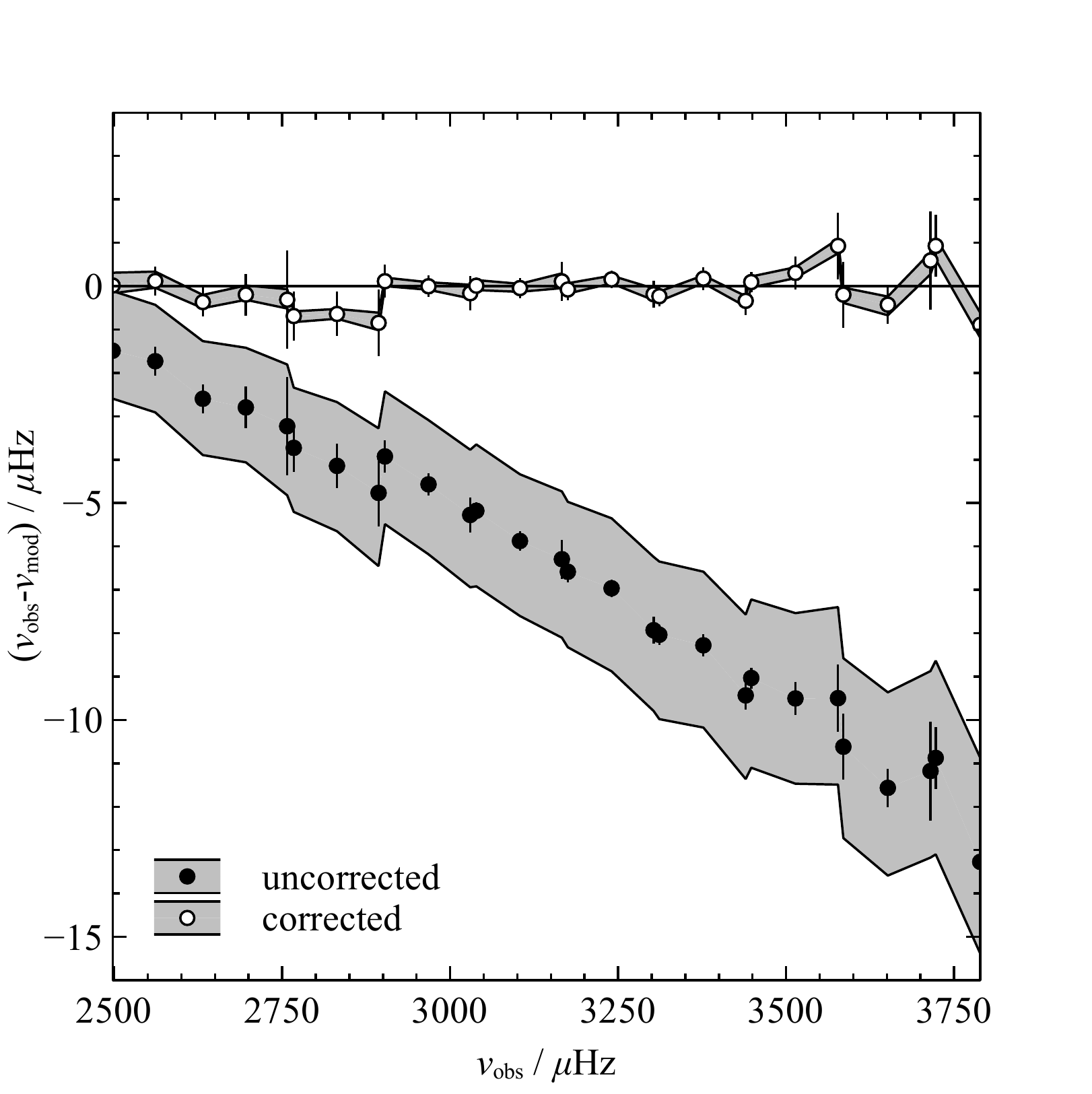}
\caption{As in Fig.~\ref{saas_cube} but with a fit made using the
  combined surface term (equation \ref{form2}).}
\label{saas_both}
\end{figure}

\begin{figure}
\includegraphics[width=85mm]{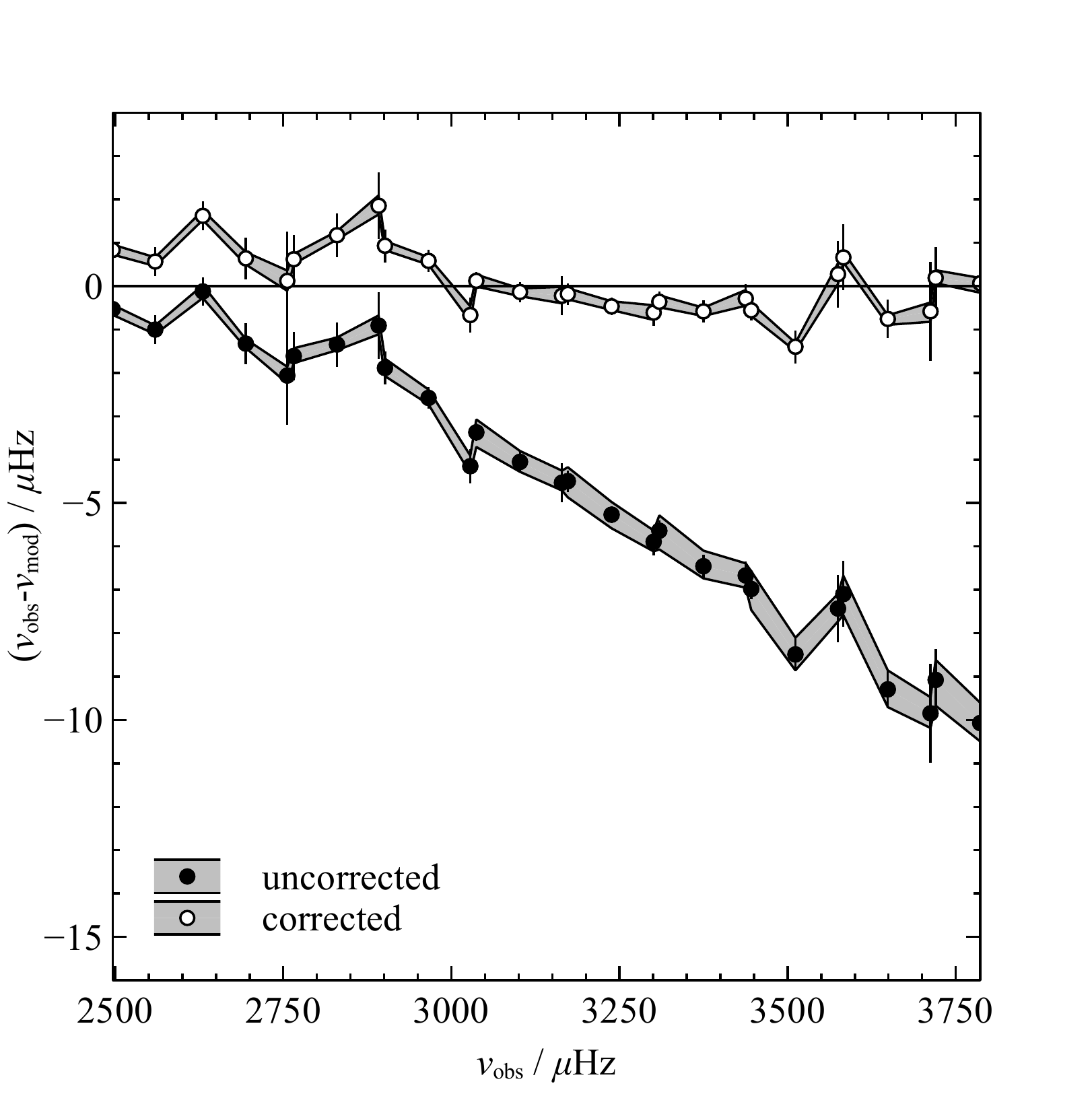}
\caption{As in Fig.~\ref{saas_cube} but with a fit made using the
  rescaled power law proposed by \citet{kbcd2008}.  Note that, as seen
  for the fits to the solar data in Figs~\ref{modelS_BiSON} and
  \ref{modelM_BiSON}, the correction is slightly overestimated at low
  frequencies.}
\label{saas_kbcd}
\end{figure}

\subsection{Solar data}

As a second test, we perform a Sun-as-a-star experiment, where we
degrade observations of the Sun to the same selection and
uncertainties as observations of HD 52265.  We fit the observations
with the cubic term (case E), the combined terms (case F), or the
scaled power law (case G), the results of which are presented in
Table~\ref{supertable}.

The fitted parameters are all consistent with our knowledge of the
Sun, although the mixing length parameter $\alpha$ is consistently
somewhat lower than the calibrated value of $1.908$.  The oscillation
frequencies tightly constrain the mechanical structure of the star and
thus the mass, radius and combinations thereof, like mean density and
surface gravity.  As noted when fitting the synthetic data, the
inverse term is not well-constrained, but its inclusion does not appear
to greatly worsen the quality of the fit.  The surface terms
themselves are consistent with the coefficients determined by fitting
the BiSON frequencies to the calibrated solar model (see
Fig.~\ref{modelM_BiSON}): $a_3=\sci{-2.13}{-7}\uHz$ for the cubic fit,
and $a_3=\sci{-2.25}{-7}\uHz$ and $a_{-1}=\sci{1.73}{-9}\uHz$ for the
combined fit.

The frequency differences for cases E, F, and G are shown in
Figs~\ref{saas_cube}, \ref{saas_both}, and \ref{saas_kbcd}.  The error
bars on the data points represent the observed uncertainties; the
shaded regions around the corrected and uncorrected differences
represent the spread in the 100 fits to independent realizations.  The
quality of the new fits to the corrected frequencies is encouraging,
given that there are no obvious remaining trends in either case.  The
power-law fit shows a marginal trend in the residual differences, with
the corrected model frequencies being too high at low frequencies.
This is consistent with the fits to solar data in
Figs~\ref{modelS_BiSON} and \ref{modelM_BiSON}.

It is clear from Figs~\ref{saas_cube} and \ref{saas_both} that
simultaneously fitting the surface effect with the cubic or combined
terms allows a greater range of models than the observed
uncertainties.  This can be seen in the fact that the envelope of
model uncertainties is broader for the uncorrected than the corrected
frequencies.  In essence, the uncertainty about the surface effects is
reflected in the uncertainties of the underlying stellar parameters
but every model is optimally corrected to match the observed
frequencies.  For example, consider case E, where only the cubic term
was used.  The uncertainty in the coefficient $a_3$ is about 10 per
cent, which naively permits an uncertainty of about $1\uHz$ for a
$10\uHz$ frequency shift, all else being equal.  In reality, the
coefficients are correlated with the underlying model parameters, so
this estimate is only approximate.  When using both terms, an even
greater range of models is allowed and the uncertainties are
correspondingly larger.

In addition, the uncertainties of the corrected model frequencies are
generally smaller than the observed uncertainties.  This follows from
the random realization process.  In the observations, the
uncertainties are uncorrelated, so the observed frequencies are
perturbed independently.  In the stellar models, the variations of the
frequencies with respect to the stellar parameters are instead tightly
correlated.  It is highly unlikely that a random realization of the
observations reproduces the correlated variations of the model
frequencies, so the underlying parameters vary relatively little from
their best-fitting values.

\section{Application to HD~52265}

HD~52265 is a metal-rich G0V planet-hosting star that was observed by
CoRoT for 117 days between November 2008 and March 2009.
\citet{ballot2011} reported spectroscopic and seismic observations
that \citet{escobar2012} used to perform a detailed analysis of the
star.  This involved the direct comparison of manually-selected models
(TGEC) and two automated fits using the Asteroseismic Modelling Portal
\citep[AMP,\footnote{\url{https://amp.phys.au.dk/}}][]{metcalfe2009}:
one using all the frequencies given by \citet{ballot2011} (fit (a))
and one omitting the lowest three frequencies, as is done here (fit
(b)).  \citet{lebreton2012} estimated the extent of inward convective
overshooting using stellar models with $t\approx2\Gyr$,
$M\approx1.25\Msun$, and $R\approx1.3\Rsun$, in reasonable agreement
with the results presented here, although they do not report
quantitative uncertainties on the model parameters.  They do estimate
the depth of the convective zone, with penetrative overshooting, to be
$R\st{cz}=0.800\pm0.004\,R_*$, where $R_*$ is the radius of the star.
This is compatible with our result of $R\st{cz}/R_*=0.803$ for the
best-fitting stellar model.
Finally, HD~52265
was modelled using the SEEK pipeline \citep{quirion2010}, as reported
in \citet{escobar2012} and \citet{gizon2013}, which returned best-fit
parameters $t=2.37\pm0.29\Gyr$, $M=1.27\pm0.03\Msun$, and
$R=1.34\pm0.02\Rsun$.  The higher mass is somewhat discrepant, and
\citet{escobar2012} argued that it represents a local, secondary,
optimum of the model parameters.  The spectroscopic observations are
listed in Table \ref{supertable}, along with model parameters for the
TGEC and AMP fits by \citet{escobar2012}, and the results presented in
this paper.
For the frequency selection, we used the reported frequencies for 10
$\ell=0$, 10 $\ell=1$ and 8 $\ell=2$ oscillation modes between $1600$
and $2550\uHz$.  We omitted the lowest three frequencies given by
\citet{ballot2011}, which are reportedly less reliable.

\begin{figure}
\includegraphics[width=85mm]{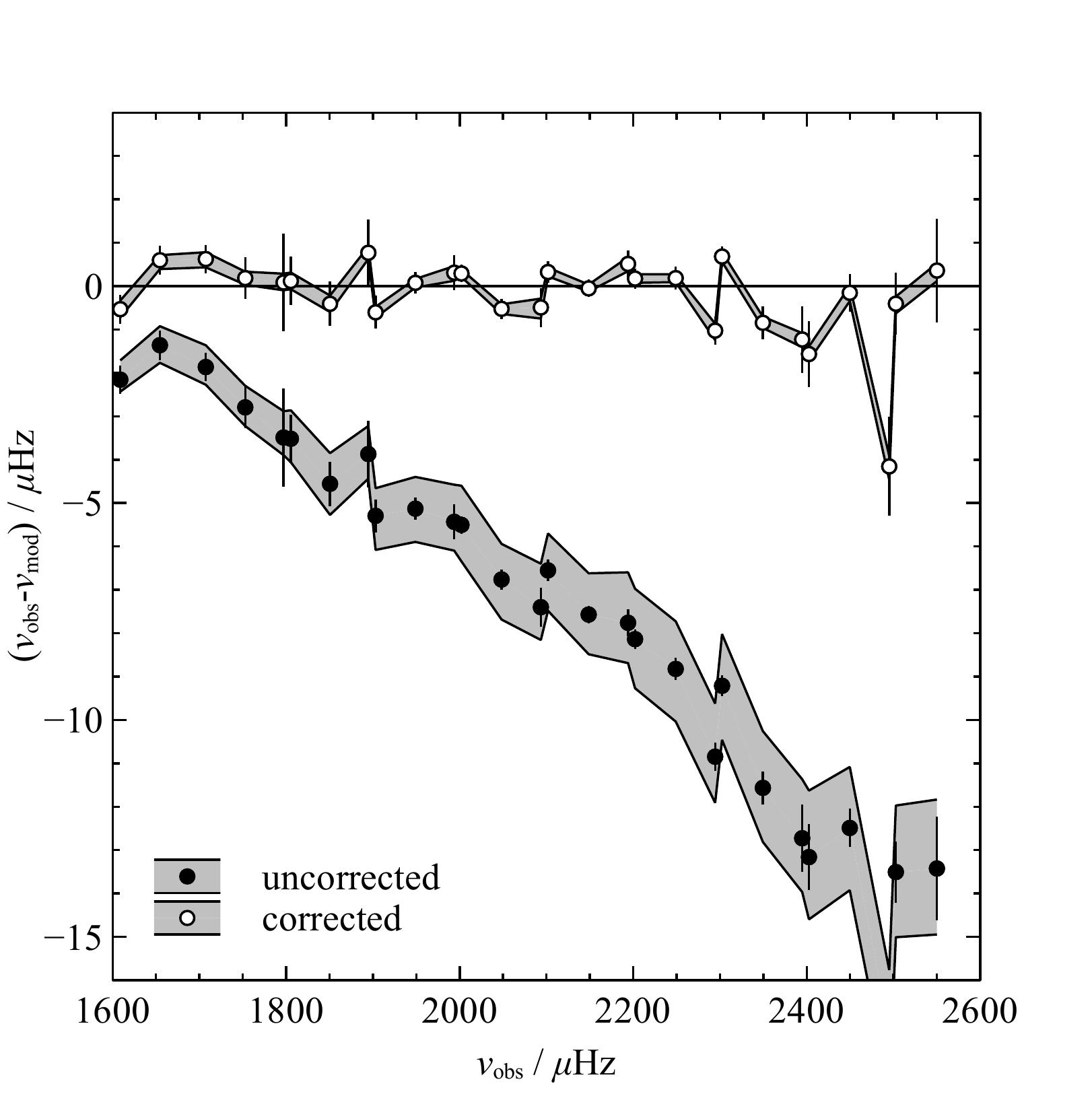}
\caption{As in Fig.~\ref{saas_cube} for the fit to HD~52265 using the
  cubic term (equation \ref{form1}).}
\label{bob_cube}
\end{figure}

The fits with the cubic, combined, and power-law surface terms (cases
H, I, and J) are included in Table~\ref{supertable}.  We have also
included the best-fit parameters reported by \citet{escobar2012} for
comparison, and our results with the cubic and combined surface terms
are entirely consistent.  Like previous models, we find that the
surface gravity is lower (though not significantly) than the
spectroscopic value.  In addition, the parameter fits made with the
combined surface term are consistent with those made with the cubic
term only.

\begin{figure}
\includegraphics[width=85mm]{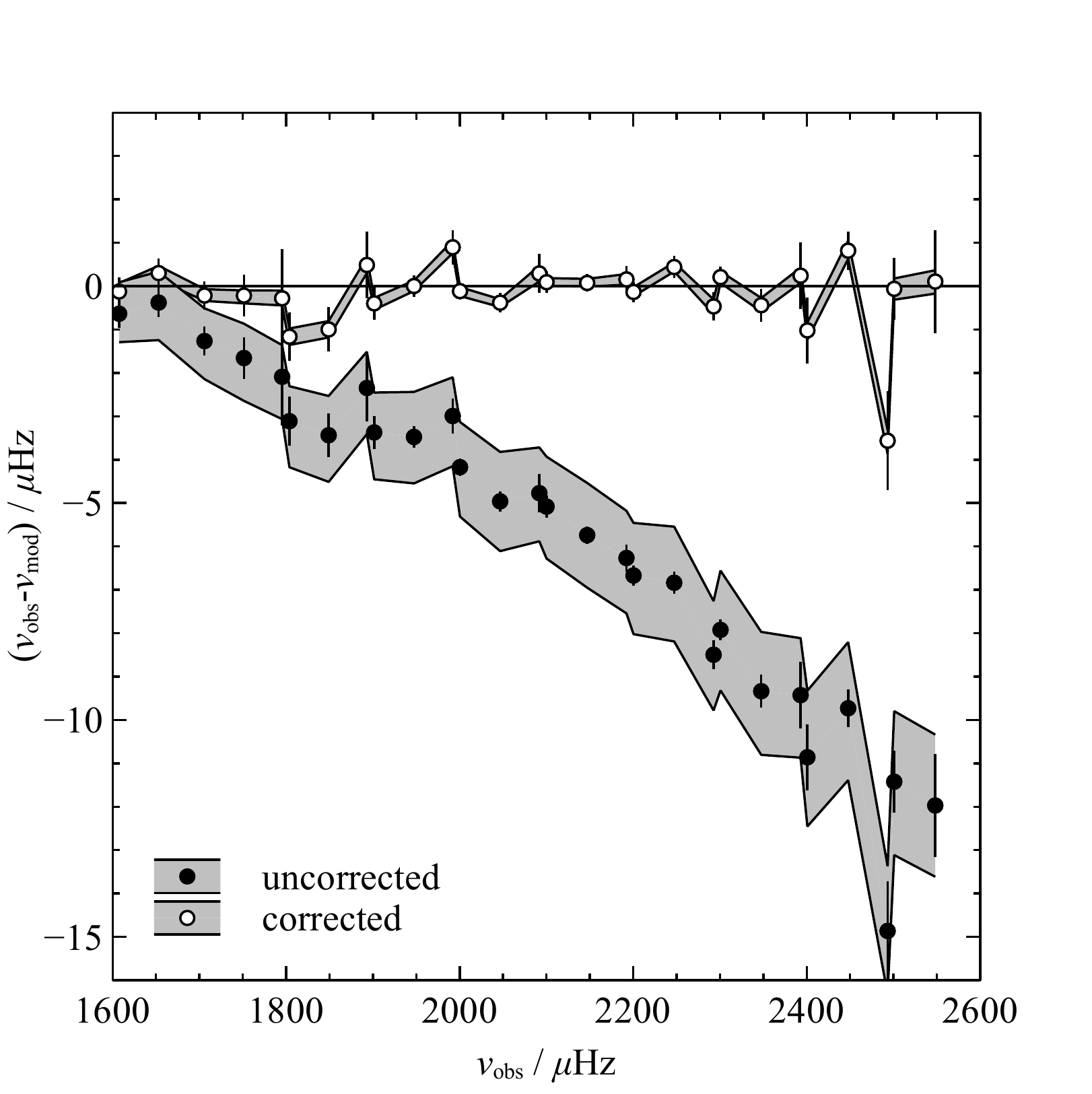}
\caption{As in Fig.~\ref{saas_cube} for the fit to HD~52265 using the
  combined term (equation \ref{form2}).}
\label{bob_both}
\end{figure}

\begin{figure}
\includegraphics[width=85mm]{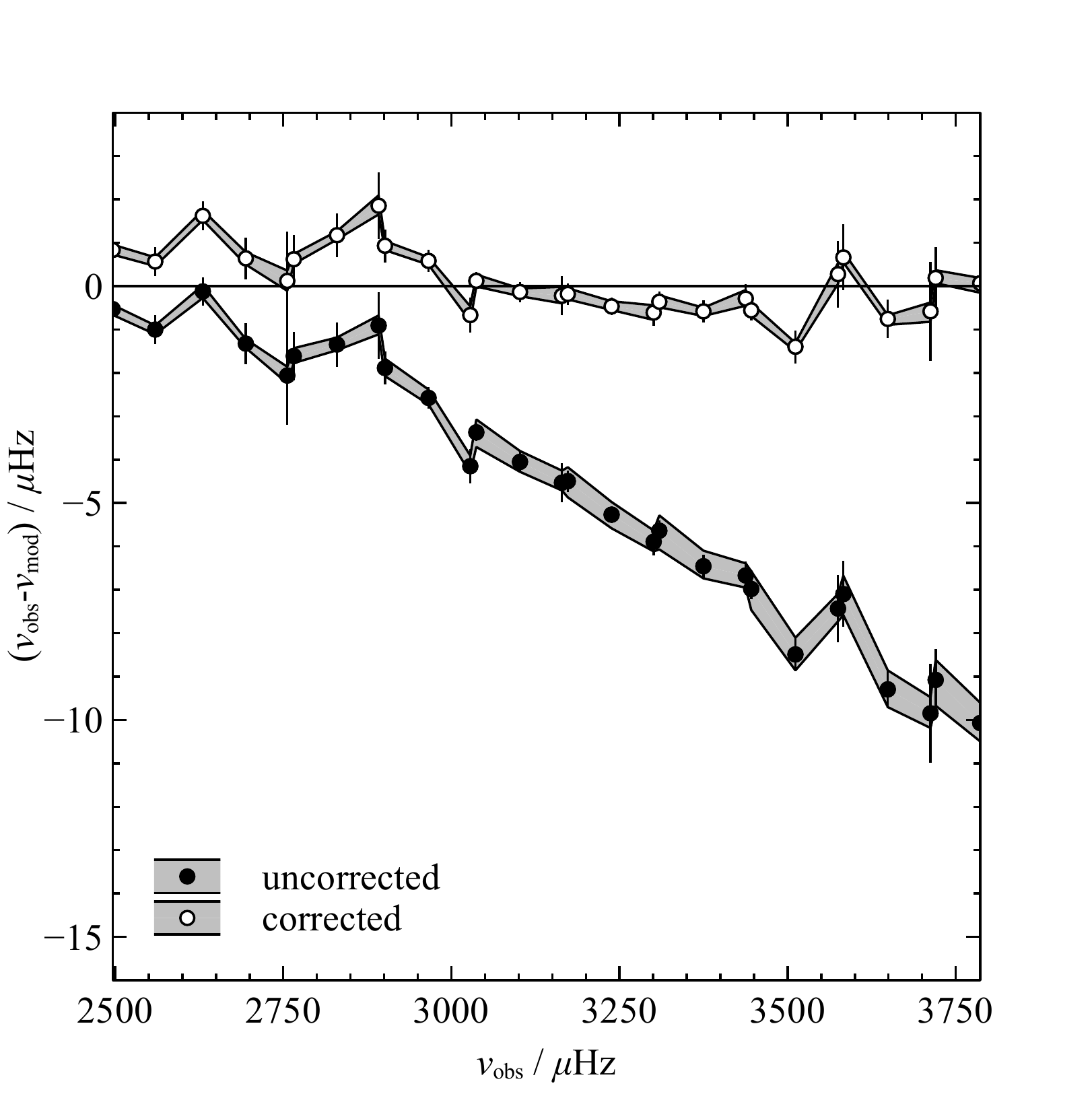}
\caption{As in Fig.~\ref{saas_cube} but for a fit to HD~52265 made
  using the rescaled power law proposed by \citet{kbcd2008}.  Again,
  the correction is slightly overestimated at low frequencies.}
\label{bob_kbcd}
\end{figure}

Figs~\ref{bob_cube}, \ref{bob_both}, and \ref{bob_kbcd} show the
frequency differences for the cubic, combined, and power-law surface
terms.  For the cubic and combined terms, there is again no obvious
residual, though the highest-frequency $\ell=2$ mode shows
a significant remaining difference.  In both cases, the overall scale
of the surface effect is similar to that of the Sun, taking a value of
roughly $7\uHz$ at $\nu\st{max}\approx2100\uHz$.  The coefficients of
the cubic term also of similar size, when the oscillation frequencies
are normalized against the acoustic cutoff frequency.  The coefficient
of the inverse term is non-zero at the $2\sigma$ level, although still
consistent with the Sun-as-a-star value because of the large
uncertainties.  As noted before, the addition of the inverse term does
not appear to improve the fit significantly, although it similarly
does no harm.  This is also presumably due to the limited number of
radial orders covered by our frequency selection.

The power-law fit again shows a slight residual trend but the overall
correction is similar.  However, the quality of the fit is
significantly worse than the cubic term.  The differences between the
corrected and observed frequencies for the best-fitting models give
$\chi^2=124$ for the scaled power-law compared with $\chi^2=31$ for
the cubic term.  The underlying model parameters deviate only within
uncertainties but we expect that they would deviate further as the
number of observed modes increases and the shape of the surface effect
deviates from a power law.  Thus, given the smaller values of $\chi^2$
in both HD~52265 and the full set of BiSON frequencies, we believe
that the cubic term produces better best-fit parameters than the
scaled power law.

\section{Discussion}

\subsection{Quality of parametrization fits}

Based on the results presented here, it appears that the proposed
parametrizations of the surface effects can adequately describe the
true surface effects, certainly enough to correctly recover the
underlying stellar model parameters.  In all cases, the price to pay
is increased uncertainties in the uncorrected frequencies, even though
the corrected frequencies have small uncertainties.  Despite this loss
of accuracy, the exploitation of the full set of frequencies, even
given a relatively meagre selection of mode frequencies, allows
precise determination of the underlying stellar model parameters.

For the degraded BiSON data, the scaled power-law correction gives
consistent results.  In this case the index of the power law has been
fit directly to the star in question, and the fit is about as good as
the new parameterizations over the observed range.  There is, however,
an uncorrected trend, which would induce greater discrepancies in the
stellar model parameters given a larger frequency range.  In the case
of HD~52265, the fit is significantly worse, although the stellar
parameters continue to be consistent.  The poor fit is not because a
power law describes the frequency differences badly, but rather
because the solar-calibrated value of the index is incorrect.  For
comparison, if we fit a power law to the frequency differences of the
best fit model with the cubic term, we find $b=4.11$, which is smaller
than the solar-calibrated value of $b=4.81$.  The use of the mode
inertia appears to capture the variation of the power law between the
two stars studied here.

In the Sun, the surface effect appears to be dominated by the cubic
term and the frequency range used here does not strongly constrain the
inverse term.  We recommend that stellar models are first fit using
only the cubic term and that the inverse term be introduced as an
additional parameter after the initial fit has been performed.  A
greater number of lower-order frequencies would better constrain the
inverse term to determine whether its contribution remains small.  In
fact, generally speaking, lower order modes constrain any surface term
better, because those modes are affected least.  Such observations are
most easily made when frequencies are determined from radial velocity
data rather than photometric data.  The proposed Stellar Observations
Network Group \citep[SONG,][]{grundahl2014}, of which one node is now
essentially operational, would be invaluable in making such
observations.

\subsection{Uncertainties}

Our uncertainties for the parameters of HD~52265 are generally smaller
than those reported by \citet{escobar2012} for the TGEC models.  Their
model parameters were derived from the mean and standard deviation of
three manually chosen best-fitting models.  The results from the AMP
are reported without uncertainties.  Thus, our smaller uncertainties
are not necessarily at odds with the TGEC fit.  The uncertainty on the
initial helium abundance $Y_i$ may seem particularly small, but it is
consistent with the uncertainty on the initial metallicity.  If we
suppose that the initial helium and metal abundances are related by
some enrichment law$Y=Y_0+(\Delta Y/\Delta Z)Z$ and that
\eq{[{\rm Fe/H}]=\log (Z/X)-\log (Z/X)_\odot\text{,}}
with some primordial helium abundance $Y_0$, then it is possible to
solve for the $Y$ given $[$Fe/H$]$.  By taking, for example,
$Y_0=0.245$ and $\Delta Y/\Delta Z=1.4$, then a change in $[$Fe/H$]_i$
from $0.27$ to $0.30$ induces a change of just $0.002$ in $Y_i$.
Thus, the uncertainties, though small, are compatible with our
knowledge of the initial metallicity, even though the helium and metal
abundance were fit independently.

The uncertainties of the fits to the synthetic data are smaller still
because the central values, about which the random realizations are
made, are perfectly consistent.  Similarly, central values of the
degraded solar data are based on the BiSON data, for which the
uncertainties are also very small.  Thus, even though we have used
greater uncertainties to randomly realize observations, the parameter
space is dominated by a deeper optimal solution.

We also note that the uncertainties reported here are purely
statistical.  They do not reflect uncertainties in the input model
physics that would be induced by changing, for example, the opacity
tables, equation of state, or atmospheric model.  This is an
interesting problem that has not yet been extensively studied for fits
using individual frequencies.  \citet{chaplin2014} analyzed several
hundred {\it Kepler} short-cadence targets using six stellar evolution
codes across 11 pipelines and concluded that the uncertainties induced
by different codes and pipelines are about 3.7, 1.7, and 16 per cent
in mass, radius, and age, when spectroscopic constraints on
$T\st{eff}$, $\log g$ and $[$Fe/H$]$ are available.  These fits were
not based on individually-resolved frequencies, which presumably
constrain models better than overall seismic properties, but they
still inform us on the differences caused by using different stellar
models.  In the future, we intend to study the systematic errors in
detail by exploiting the modular design of MESA to isolate the effects
of individual choices of input physics.

\subsection{Potential shortcomings}

Despite performing well on the targets studied here, the method does
contain several potential flaws.  First, as is always the case with
optimization problems, we cannot guarantee that we have found the
global best-fit solution.  To some extent, the repeated random
realizations mitigate this problem because if a much better minimum
were available, we might expect at least a few of the random
realizations to converge there.  Indeed, this is the case in HD~52265
where, in cases H and I, about 15 and 3 models settled on a secondary
minimum with a mass of roughly $1.30\Msun$, close to the value
identified by \citet{escobar2012}.  The asymmetry of the uncertainties
in Table \ref{supertable} captures the existence of this alternative
local optimum.  An initial grid-based search could also rule out the
existence of better minima in distinctly different regions of
parameter space.  The greater problem is that the introduction of one
or two new free parameters, which are simultaneously optimized during
the fit and currently not limited in scale or sign, might introduce
new local minima that confuse the optimization process.  We propose to
mitigate this by choosing initial guesses that weight non-seismic
observations more heavily, though the final sample of fits should
continue to be performed without artificially weighting the
observables.

Second, there is an additional implicit free parameter: the depth of
the normalization of the mode inertiae, which controls the depth of
the perturbation that produces the surface effect.  Here, we have used
the radius of the photosphere.  This is a reasonable choice and fits
the Sun well but there is no guarantee that this will be the case for
stars significantly different from the Sun.  \citet{rosenthal1999}
computed the frequency differences between a solar model and a
detailed simulation of the Sun's surface convection zone and found
that the surface effect is largely (but not entirely) produced by the
suspension of the envelope by turbulent pressure, which peaks in a
narrow region that is essentially at the photosphere.  Recent
simulations \citep[e.g.][]{beeck2013,magic2013,trampedach2013} should
allow similar studies to be carried out for a range of stellar surface
properties and allow some forward computation of the expected location
and scale of the surface correction.  Presently, it seems reasonable
to continue to normalize the mode inertiae at the photosphere.

Third, our assessment of the parametrizations is specific to the
traditional mixing-length theory and adiabatic pulsations.  Though not
widely implemented, other one-dimensional descriptions of convection
exist \citep[e.g. \emph{full spectrum turbulence},][]{canuto1991} and
these predict different properties for the superadiabatic layers near
the surfaces of Sun-like stars.  Since this region introduces some
component of the surface effect, the parametrizations presented here
may no longer fit so well.  Similarly, non-adiabatic pulsation
frequencies exhibit a kind of surface effect, which also may or may
not be described precisely by the parametrizations.

\section{Conclusion}

We have presented two parametrizations of the known systematic
difference between modelled and observed stellar oscillation
frequencies induced by poor-modelling of the near-surface layers.  The
first contains a term proportional to $\nu^3/\mathcal{I}$ (the
\emph{cubic} term, equation \ref{form1}) and the second an additional
term proportional to $\nu^{-1}/\mathcal{I}$ (the \emph{inverse} term,
equation \ref{form2}).  The second parametrization corrects most of
the frequency differences between two calibrated solar models and
observed solar oscillation frequencies, in the sense that no obvious
trends remain.  Both parametrizations fit the frequency differences
significantly better than power laws.

We implemented a simultaneous fit to the surface effects in MESA and
constructed a model-fitting pipeline that fits stellar models to 100
random realizations of the observable data.  By testing the new
pipeline on data for a synthetic target, the Sun, and the CoRoT target
HD~52265, we have shown that these parametrizations give unbiased
results that are compatible with independent measurements and model
fits.  We compared our results with calculations made with the same
pipeline, but using the scaled power-law proposed by \citet{kbcd2008}.
In the case of the Sun, both methods perform similarly well, but for
HD~52265, the power-law fit produces a significantly worse fit to the
frequencies but consistent stellar parameters.  Thus, the cubic term
alone is arguably an improvement over a scaled power law.

Generally, the cubic term dominates the surface correction.  The
contribution of the inverse term is clear when comparing observations
to calibrated solar models, but poorly-constrained by the small range
of radial orders that we have used in the fitting pipeline.  In all
cases, the stellar model parameters are consistent, irrelevant of
which surface term is used, and the overall scale of the surface term
is similar in the Sun and HD~52265.

The simultaneous fit potentially introduces new local minima to the
models' parameter space and we recommend that initial guesses be found
by weighting complementary non-seismic observations more strongly.
Given a reasonable initial guess, however, we have found that the
pipeline converges on a reasonable solution even though the magnitudes
and signs of the coefficients are not constrained.  In upcoming work,
we will apply our pipeline to more stars, covering a greater range of
parameters, to ensure that the parametrizations are generally useful
in allowing us to exploit present and future observations of resolved
oscillation modes.

\begin{acknowledgements}
  WHB and LG acknowledge research funding by Deutsche
  Forschungsgemeinschaft (DFG) under grant SFB 963/1 ``Astrophysical
  flow instabilities and turbulence'' (Project A18).  
  LG also acknowledges support from EU FP7 Collaborative Project
  ``Exploitation of Space Data for Innovative Helio- and
  Asteroseismology" (SPACEINN).  Part of this work was completed
  during a visit to the Aarhus Stellar Astrophysics Centre.
\end{acknowledgements}


\end{document}